\documentclass{emulateapj}
\usepackage{graphicx}
\usepackage{epstopdf}

\newcommand\gtsim{\mathrel{\lower0.6ex\hbox{$\buildrel {\textstyle >}\over {\scriptstyle \sim}$}}}
\newcommand\ltsim{\mathrel{\lower0.6ex\hbox{$\buildrel {\textstyle <}\over {\scriptstyle \sim}$}}}
\newcommand\Lo{$\mathcal{L}_0$}
\newcommand\ro{$r_0$}
\newcommand\vk{von K\'{a}rm\'{a}n }
\def\mathbi#1{\textbf{\em #1}}

\newcommand\rse{$r_s = \mathrm{FWHM_{DIMM}}-\mathrm{FWHM_{science}}$}
\usepackage{aas_macros}
\usepackage{url}
\usepackage[footnotesize]{subfigure}
\shortauthors{Floyd et al.}

\slugcomment{Accepted for publication in PASP}
\shorttitle{Image Quality at Las Campanas}
\shortauthors{Floyd et al.}

\begin{document}
\title{Seeing, Wind and Outer Scale Effects on Image Quality at the Magellan Telescopes}
\author{David J. E. Floyd\altaffilmark{1,2} \email{dfloyd@unimelb.edu.au}
Jo Thomas-Osip\altaffilmark{2}, 
Gabriel Prieto\altaffilmark{2}
}
\altaffiltext{1}{AAO/OCIW Magellan Fellow. 
Current address: School of Physics, The University of Melbourne, VIC, 3010, Australia}
\altaffiltext{2}{GMT Organsitation, Las Campanas Observatory, Casilla 601, Colina El  Pino, La Serena, Chile}

\begin{abstract}
We present an analysis of the science image quality obtained on the twin 6.5~m Magellan telescopes over a 1.5 year period, using images of $\sim10^5$ stars. We find that the telescopes generally obtain significantly better image quality than the DIMM-measured seeing. This is qualitatively consistent with expectations for large telescopes, where the wavefront outer scale of the turbulence spectrum plays a significant role. However, the dominant effect is found to be wind speed with Magellan outperforming the DIMMs most markedly when the wind is strongest. 
Excluding data taken during strong wind conditions ($>10$~m~s$^{-1}$), we find that the Magellan telescopes still significantly outperform the DIMM seeing, and we estimate the site to have $\mathcal{L}_0\sim25$~m on average. 
We also report on the first detection of a negative bias in DIMM data. This is found to occur, as predicted, when the DIMM is affected by certain optical aberrations and the turbulence profile is dominated by the upper layers of the atmosphere.
\end{abstract}

\keywords{Astronomical Phenomena and Seeing -- Astronomical Instrumentation -- Astronomical Techniques -- Data Analysis and Techniques}

\clearpage
\section{Introduction}
\label{sec-intro}
Much time and money is invested in finding optimal sites for the next generation of big optical-IR telescopes and it is of interest to learn the detailed atmospheric properties of major observatory sites since the seeing is a major limitation to the resolution and signal to noise ratio that can be achieved. 
It is well known that large aperture telescopes frequently perform better than we would naively expect from the simple seeing measurements at major observatories (e.g.~\citealt{martin+98, wilson+99, tok+07,salmon+09,els+09}) in terms of image full width at half maximum (FWHM).
Most astronomers consider the ``seeing'' to be synonymous with the FWHM of a stellar image on their detector, but this assessment is formally incorrect. Seeing is a simple measurement of Fried's parameter, $r_0$~\citep{fried65}, determined using differential image motion monitors (DIMMs -- see~\citealt{DIMM} and references therein) but there are other important parameters governing imaging quality, such as the characteristic time constant, the isoplanatic angle and the wavefront outer scale, \Lo.
These can play important roles in different regimes.

\subsection{Background: Turbulence, seeing and outer scale}
Using a single parameter to determine image FWHM implicitly assumes that energy is injected into the atmosphere at infinite scale following the Kolmogorov turbulence model~\citep{tatarskii69,kolm41}, an approximation that is valid only for small telescopes.
In the Kolmogorov model, energy is gradually transferred to smaller and smaller scales until it is dissipated through the viscosity of the air on scales of $\sim$ mm -- cm (the inner scale, $l_0$). The result is local variations in atmospheric temperature and density, and thus in refractive index, $n$. 
In reality the atmosphere continually has energy injected into it through heating, convection currents and wind shear on scales $L_0\sim10$~m, which we can introduce as a cutoff in the turbulence spectrum or ``outer scale of turbulence'', \Lo. This is known as the \vk turbulence model (see~\citealt{borgnino90,ziad+00}). 

In practice atmospheric turbulence results in small amounts of image motion. The standard device for measuring this motion is the DIMM. A DIMM measures the variance of the differential 
image motion, $\sigma_d^2$, of two images of the same star, observed through separate apertures in the entrance pupil of a small telescope. 
This gives a measurement of \ro\ for a sub-aperture of diameter $D$ and wavelength $\lambda$:
$\sigma_d^2 \propto \lambda r_0^{-5/3} D^{-1/3}$.
Thus the FWHM $\epsilon_0$ of the long-exposure seeing-limited point spread function in large telescopes is computed within the Kolmogorov turbulence model as $\epsilon_0 = 0.98 \lambda/r_0 \propto ( D/\lambda )^{1/5} (\sigma_d^2)^{3/5}$.

\begin{figure}[htbp]
\begin{center}
\plotone{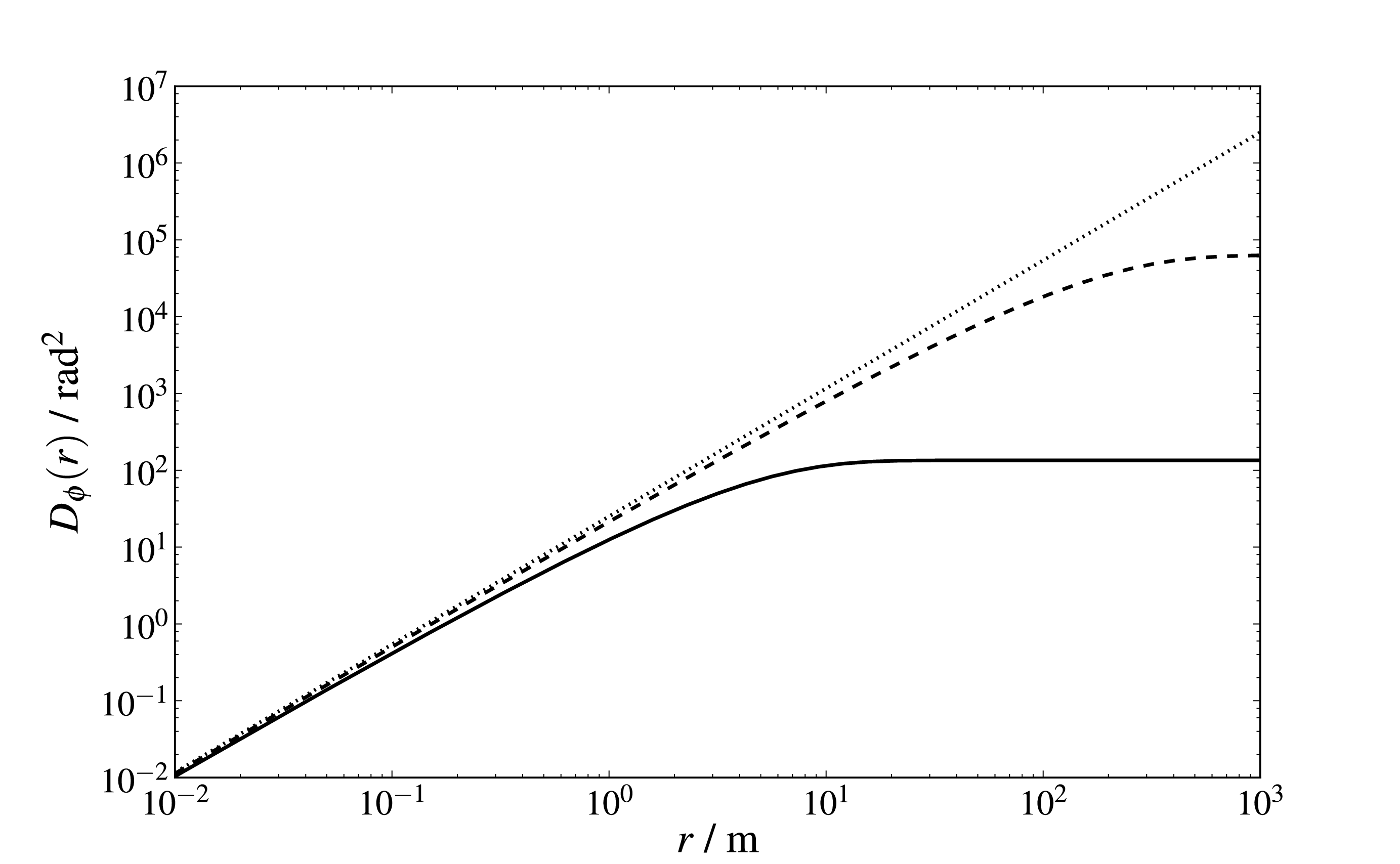}
\caption{\label{fig-strucfun} The \vk phase structure function, $D_\phi(r)$ (solid line $\mathcal{L}_0=25$~m, dashed line $\mathcal{L}_0=1000$~m) compared to the Kolmogorov phase structure function (dotted line), all for $r_0=0.2$~m. Unlike the Kolmogorov structure function which continues to grow with increasing $r$, the \vk structure function saturates at $r=$\Lo. The outer scale effect becomes noticeable at scales significantly smaller than \Lo.}
\end{center}
\end{figure}

From a turbulence spectrum it is possible to derive a statistical structure function $F_\phi(\mathbi{r})$ which describes the phase decorrelation between any pair of points in the atmosphere, separated by $\mathbi{r}$. 
In the Kolmogorov model, the phase structure function (the mean square differential phase error -- see~\citealt{roddier81} for a useful derivation) grows larger with distance of separation, 
$D_\phi(\mathbi{r}) \propto (|\mathbi{r}|/r_0)^{5/3}$  -- see dotted line in Fig.~\ref{fig-strucfun}. 
But in the \vk model the structure function becomes more complicated~\citep{ConRonch72} and 
grows rapidly with $r$ up to $r = \mathcal{L}_0$ where it saturates -- see solid line in Fig.~\ref{fig-strucfun}: 
Beyond \Lo, $D_\phi(\mathbi{r})$ 
no longer increases. The outer scale effect is detectable in telescopes that are an appreciable fraction of the size of \Lo, where it begins to deviate from the Kolmogorov case. 

From the structure function we can derive the optical transfer function (OTF) of a large telescope, $T_0(\mathbi{f}) = \exp[-0.5 D_\phi (\lambda\mathbi{f})]$,
where $\mathbi{f}$ is the spatial frequency in inverse radians. 
The corresponding Point Spread Function (PSF) is found by taking the Fourier Transform of the OTF. 
\citet{tok02} obtained the PSF for the \vk case by renormalizing the OTF. 
The result is an approximation of the seeing-limited PSF of a large telescope, with FWHM $\epsilon_{\mathrm vK}$ approximated by:
\begin{equation}
\label{eq:vk}
\left( \frac{\epsilon_{\mathrm vK}}{\epsilon_0} \right)^{2} \approx 1 - 2.183 \left( \frac{r_0}{\mathcal{L}_0} \right)^{0.356}
\end{equation}
which we use later in this paper to estimate an average value of \Lo\ at the Las Campanas site.
The effect of introducing an outer scale to the phase structure function is to decrease the FWHM of the point spread function (PSF) of a ``large'' telescope. 


\section{Image quality measurements at the Magellan telescopes}

The twin 6.5~m Magellan Telescopes (individually named Clay and Baade) were built and are operated by a consortium consisting of the Carnegie Institution of Washington, Harvard University, MIT, the University of Michigan, and the University of Arizona. They are situated on Cerro Manqui within the Las Campanas Observatory site at (29$^\circ$00\arcmin54\arcsec S~70$^\circ$41\arcmin32\arcsec W) on the southern edge of the Atacama desert in Chile. Cerro Manqui sits at 2,450~m, slightly lower than Las Campanas peak itself at 2,551~m. 
The telescopes are described in detail by~\citet{schec03}, and the Magellan optical design including its active optics Shack-Hartmann (S-H) system is further discussed by~\citet{schec94}. 
Optical seeing down to a FWHM$\sim0\farcs25$ have been recorded~\citep{schec03}.

In 2007 we began monitoring the image quality on all Magellan telescope science images with $\ge30$~s exposure time. 
The aim was to see how well they do compared to the nominal seeing conditions and to help diagnose misalignments in the optical path.
Furthermore, Las Campanas Observatory (LCO) is the chosen location for the Giant Magellan Telescope (GMT) for which site testing began in 2005~\citep{CoDR:GMT,jet+08}. The site testing for GMT provides us with a wealth of tools for determining the precise seeing, atmospheric turbulence and meteorological conditions  -- see section~\ref{sec-GMTtest}. The GMT site testing program made use of four DIMMs around the Las Campanas site including one for reference at the Magellan Telescope site on Cerro Manqui.

In this paper we take the image-wide average full width at half maximum (FWHM) of Magellan science-camera stellar images using a program called ``DIPSF'' (as described in section~\ref{sec:anal}) and compare them with the contemporaneous DIMM (see section~\ref{sec-GMTtest}) and Magellan guide camera FWHM measurements of the seeing. We then consider simultaneously taken atmospheric turbulence and meteorological data, using the GMT Multi-Aperture Scintillation Sensor (MASS) and weather station. 

The Magellan guide cameras use an RG610 filter, with a mean wavelength of 765.2~nm. 
DIMM measurements are calculated at 500~nm. 
The guide camera and DIPSF FWHM measurements in this paper are corrected to airmass 1 and to a wavelength of 0.5~$\mu$m following the well know $r_0 \propto  \lambda^{6/5}$ ($\epsilon_0 \propto \lambda^{-1/5}$) and $\epsilon_0 \propto \mathrm{airmass}^{3/5}$ relations.
The guide camera measurements do not necessarily reflect the actual seeing at the telescopes since the guider optics are not capable of producing images better than 0\farcs4 arcsec (FWHM). The guide stars are several arc minutes off-axis and are uncorrected for aberrations in the telescope optics.

\subsection{Image Analysis Software}
\label{sec:anal}
We use a source detection and fitting program known as DIPSF, which is based on DoPhot~\citep{schechter+93}. 
DIPSF reads in an image and bitmaps it to form a low dynamic range image in memory, which is then analyzed for sources. 
Only the most basic data reduction steps are performed on the image: Bias subtraction is performed using the overscan regions, and static bad pixel masks  (computed during engineering runs each month) are applied, including any vignetted regions of the detector. 
We stress that the raw data remains untouched -- the data is read in and all processing is performed in memory. 
Sources are are fit with a 2-dimensional gaussian and are classified as stellar, cosmic ray, double or extended according to the resulting fit parameters.
The contemporaneous guide camera image FWHM is used as a guess for the first fit, and for reference in decision making about the nature of a source. Each star detected on a given frame is recorded in an output file, and several high level statistical measures of the average point spread function across the image are stored to a database.

Magellan data are proprietary and are stored for just one month in case of corruption of a dataset that an astronomer takes home. Therefore each image is analyzed the day after it was taken. At present DIPSF is run by the instrument specialists on each data directory from a night's observing. 
Only direct images that have exposure times $>30$~s are processed in order to analyze seeing-averaged images. 
We run DIPSF on data from LDSS3 (on Clay), MagIC and IMACS (on Baade) -- see~\citet{maginst08, osip+04} for more information on the instruments themselves. There is also a mode for reading MegaCam data. 

\begin{figure}[htbp]
\begin{center}
\plotone{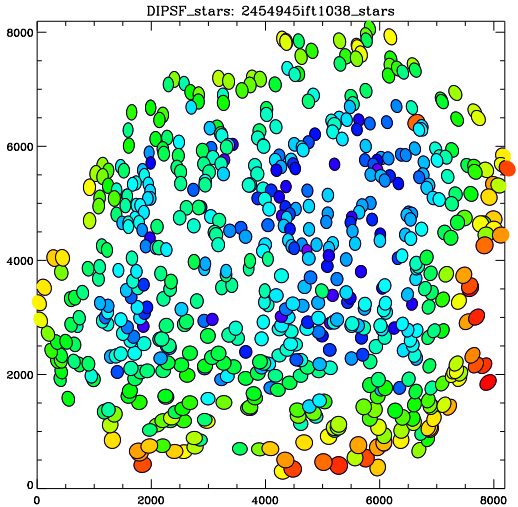}
\caption{\label{fig-stars} Example DIPSF output showing all stars detected on a crowded IMACS f/2 image from the night of 2009 April 23rd. The ellipses denote each stellar PSF and are shown with true ellipticity and position angle. Ellipse size is scaled according to the FWHM of the stellar image, and each ellipse is color-coded according to its relative size. On this occasion the mean image FWHM is $0\farcs504\pm0\farcs014$. Axis labels show pixel number measured across the IMACS mosaic.}
\end{center}
\end{figure}


The code is written in Fortran and is called from a Python script.
Output includes list of all sources identified ({\sc \_star} files), along with their properties: position, size, ellipticity, position angle, flux.
Two high level output files contain the average PSF characteristics over each chip (for multi-chip instruments) and over the entire frame ({\sc \_chip} and {\sc \_frame} files respectively).
After running, the python script posts the contents of the frame-averaged results to an SQL database. This can then be queried along with the meteorological and guide camera data from any given night or nights. The individual {\sc \_star} output files are stored on a server. 
The contents of an example IMACS f/2 {\sc \_star} file are plotted in Fig.~\ref{fig-stars} to highlight the stability of the PSF across the entire half-degree field of view. Stellar images are shown to scale with their true ellipticities and position angles and are color-coded according to FWHM. In this image the mean image FWHM is $0\farcs504\pm0\farcs014$. 
The software can also be run in a ``diagnostic'' mode which runs on {\em all} images, the results of which are simply stored locally, not posted to the database nor server. This is useful for analysis of engineering data, such as pinhole masks and out of focus ``donut'' images, and may be used to detect and diagnose defects in the telescope optical path, and potentially to supply Shack-Hartmann like information to the telescope mirror actuators. 

The program was run from May 2007 on occasional basis, and on all data from the start of 2008 to the present. 
We thus have a database containing the frame-averaged PSF FWHM for every $>30$~s LDSS3, IMACS and MagIC image taken on the Magellan telescopes since January 2008, and a few images from earlier.
The software continues to be run on a regular basis, but the DIMM's are now infrequently operated since the GMT site testing program is complete.

\subsection{GMT site testing: DIMM seeing, meteorology and turbulence measurements}
\label{sec-GMTtest}
An extensive site testing program has identified Cerro Las Campanas as the future GMT site~\citep{jet+10}.  
Four sites within the LCO property were studied.  Of these, the site of the Magellan telescopes, Cerro Manqui, was used for reference purposes only.  The seeing at all four sites was monitored through the use of identical DIMM systems in 7~m towers.   The DIMM method is based on the relationship between the FWHM of a long exposure image in a large telescope to the variances in the differences in the motion of two images of the same star through the use of Kolmogorov theory.  The two images are created by placing a mask with two sub-apertures containing prisms at the entrance of the optical tube.  The GMT instruments 
are based on the CTIO RoboDIMM\footnote{\url{http://www.ctio.noao.edu/telescopes/dimm/dimm.html}} but have several improvements. Image quality has been improved by using two thinner prisms as opposed to one thicker prism and an open aperture. Following a technique developed by the Thirty Metre Telescope (TMT) project, the Carnegie DIMM (CDIMM) software uses a drift scan readout mode which allows for many more image motion measurements to be made per minute and thus improved statistics\footnote{\url{http://users.obs.carnegiescience.edu/birk/CDIMM/}}. 
The Manqui DIMM is located approximately 20~m northwest of the Clay dome, almost exactly abeam of the Magellan telescopes with respect to the prevailing northeasterly wind. 
The Manqui peak itself is approximately 3.8~km at bearing 345$^\circ$ from Campanas peak.

The DIMM measurements are filtered to remove poor quality data using three criteria:  focus, read noise, and the number of measurements taken during one minute.  We measure focus using the mean separation of the two images during one minute of measurements.  Nominal focus is determined on a night with good seeing by varying the focus and maximizing the Strehl ratio.  The CDIMM software is designed to keep the mean separation within a range of 1 pixel on either side of the nominal focus.  There are however some measurements throughout the night (but especially at the beginning of the night) that need to be removed because the separation is outside the nominal range for a short time while the instrument works to correct it. The filters for read noise and the number of measurements allow for the removal of suspect quality data due to clouds or tracking errors.

In addition to seeing, many other parameters were also studied as a part of the GMT site testing program.  Some of these will also be used here in order to understand differences seen between the Magellan telescope (DIPSF) FWHM and the DIMM seeing FWHM.  Meteorological characteristics (pressure, temperature, humidity, and wind) were collected using weather stations manufactured by Davis Instruments Corp and mounted on 10~m towers.  Turbulence in the atmosphere above 500~m is being monitored by a MASS~\citep{kornilov+03}.  
The spatial scale of the scintillation variation depends on the distance to the layer in which the turbulence giving rise to the wave front phase disturbance exists. Thus, the turbulence profile, denoted $C_n^2(z)~dz$, at a small number of discrete layers ($z=0.5$, 1, 2, 4, 8, and 16~km) can be restored by fitting a model to the differences between the scintillation indices within four concentric
apertures.  In addition to the turbulence profile, the MASS also measures free atmospheric seeing (essentially the integral of the turbulence profile).  The difference between the total DIMM and the free atmosphere MASS turbulent integrals is a measure of the portion of the total seeing contributed by a ground layer below 500 m~\citep{tok_korn07}.

The MASS data have been reprocessed using a new version of Turbina~\citep{kornilovshatsky05} to correct for the effect of strong scintillation under poor seeing conditions~\citep{tok06}.  Strong scintillation affects the turbulence profile and free atmosphere seeing by over-estimating turbulence and spreading turbulence to lower levels. The sensitivity of the possible errors due to incorrectly determined parameters (such as the instrument magnification or the non-Poisson and non-linearity characteristics of the detectors) lead to a maximum bias on the order $\pm$0\farcs05 ~\citep{jet+08} similar to that found by~\citet{els+08}.
The seeing, $\epsilon$, contribution by the ground layer (defined here as lower than 500~m) can be inferred from the MASS and DIMM measurements in the following manner:
\[
\epsilon_{\mathrm GL} = \left| \epsilon_{\mathrm DIMM}^{5/3} - \epsilon_{\mathrm MASS}^{5/3} \right|^{3/5}
\]

The absolute value is used only in computing statistical distributions.  This need arises due to that fact that some small percentage (2--20\%) of matched MASS and DIMM measurements result in what could be termed a negative ground layer seeing values in the sense that the MASS measurement is larger than the DIMM measurement.  This is, of course, not physically possible but is a result of non-zero errors and the subtraction of quantities of similar magnitudes.  The range in the percentage of negative ground layer seeing arises when considering selections of data filtered for scintillation index in the smallest MASS aperture.  The correction made in the MASS reprocessing for strong scintillation is not perfect and this effect can be seen by examining the percentage of negative ground layer found in samples with varying scintillation indices.  The sample presented in this analysis is not filtered for scintillation index since this would introduce a bias towards better seeing.  But in this case, one must realize that the ground layer as determined from data that has not been filtered for scintillation index will be less accurate.

\subsection{DIMM bias}
\label{sec-dimmbias}
We identified a problem in the DIMM data, starting in January 2009. The Meade telescope primary mirror developed a non-repeatable movement during focus changes which caused the collimation to deteriorate.  This was discovered in the consequential deterioration of the DIMM Strehl ratios.  We therefore restrict this paper to observations taken between July 2007 and the end of December 2008.
The bias and its effects on the data taken in 2009 are discussed further in section~\ref{sec-09}.

\begin{figure*}[htbp]
\begin{center}
\plottwo{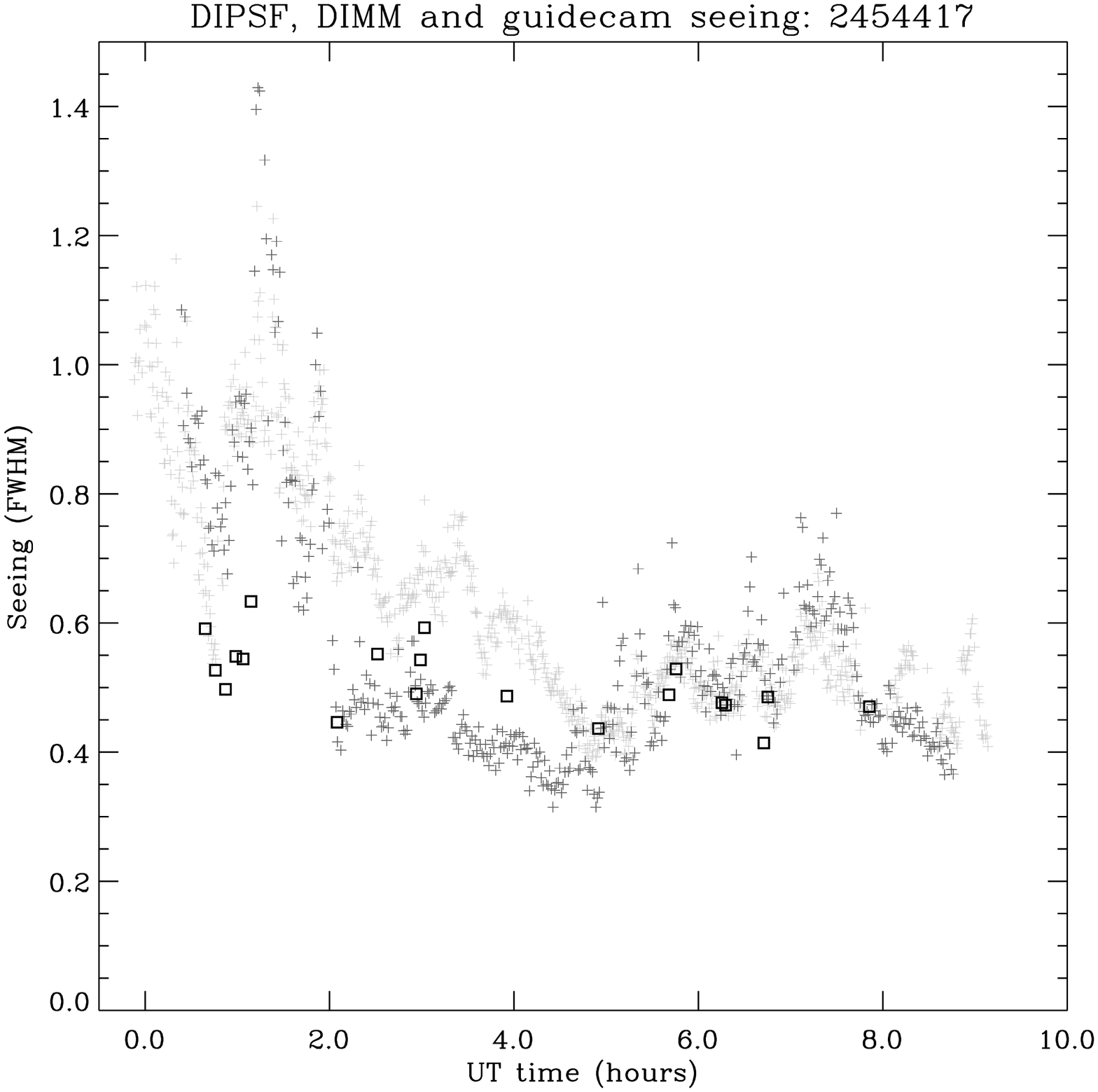}{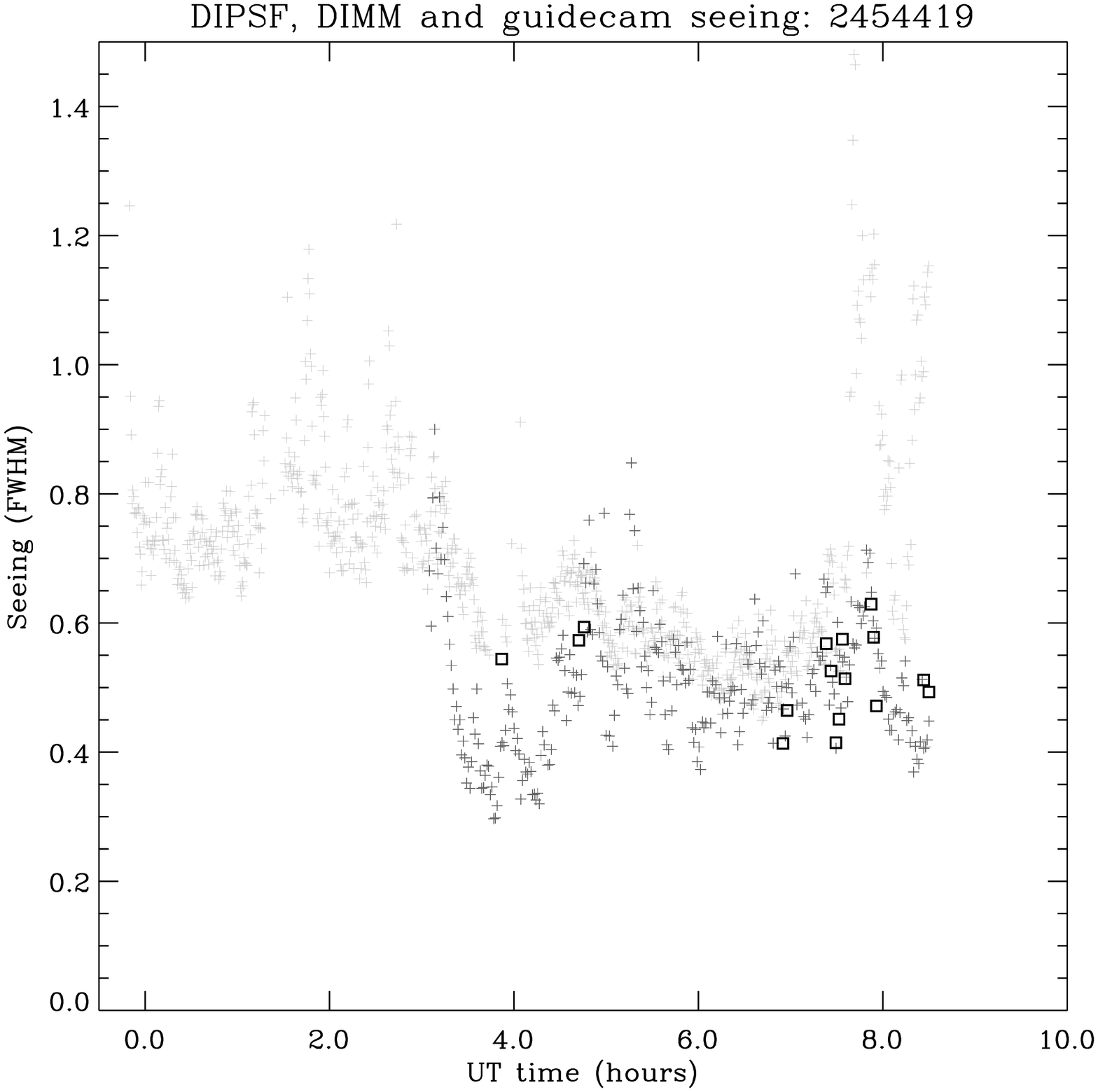}

\plottwo{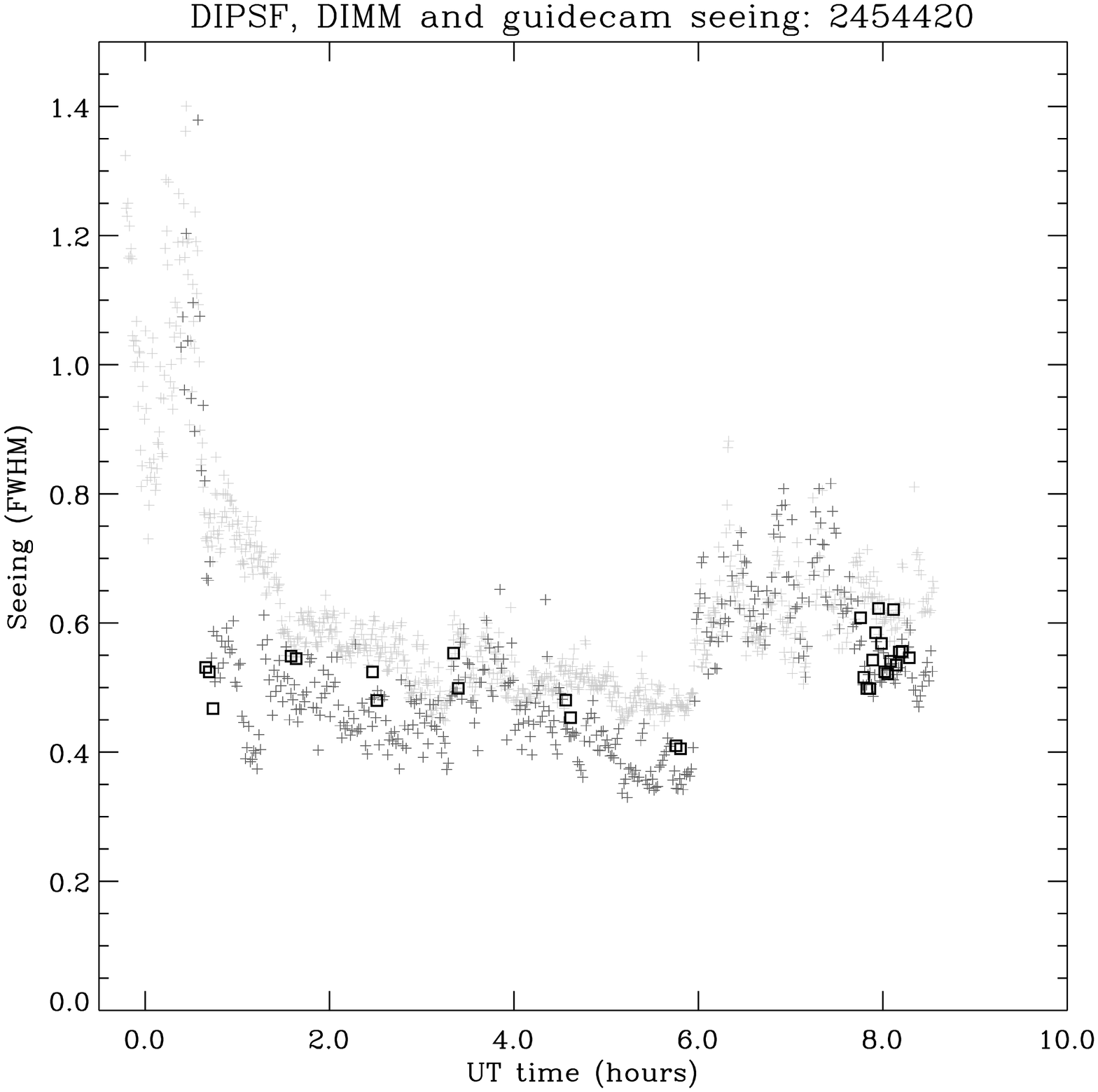}{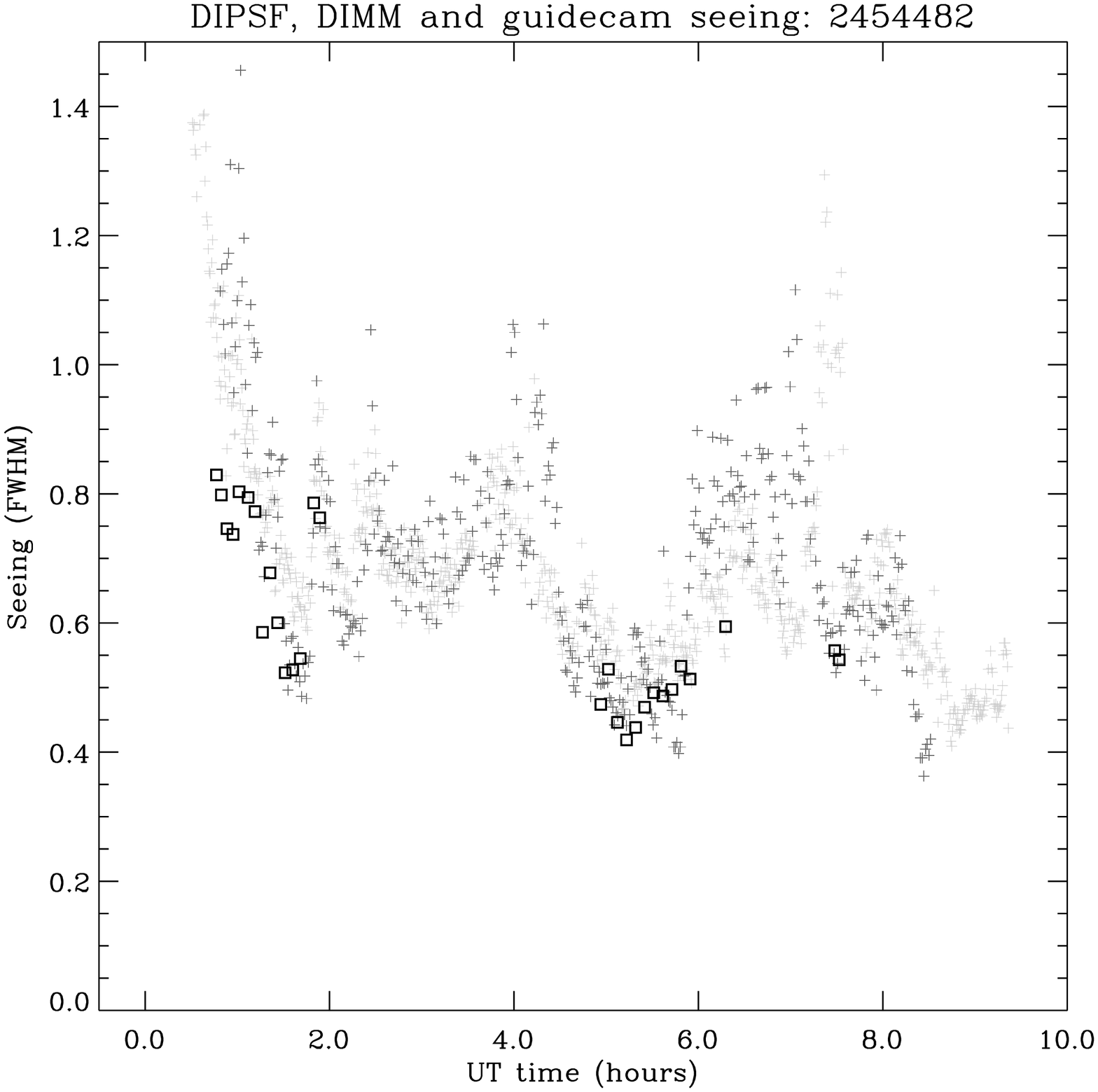}
\caption{\label{fig-time} Examples of contemporaneous DIPSF (black squares), DIMM (dark grey crosses) and guide camera (light grey crosses) seeing measurements for (ut-dates) 2007-11-12, 2007-11-14, 2007-11-15 and 2008-01-16. 
}
\end{center}
\end{figure*}

\section{The DIPSF dataset and matching to the DIMMs}
\label{sec-dataset}

We have cut out the 2009 data due to the DIMM problem mentioned in section~\ref{sec-dimmbias}. 
DIPSF data was obtained on 230 individual nights over the period July 2007 -- December 2008. 
We have identified 70 nights during this period on which there were problems with the DIPSF data, due to source detection or other software errors, or out-of-focus observations. 
Fits to $\sim7\times10^5$ stellar images were performed during this time, yielding 5617 frame-averaged FWHM datapoints with an overall average of 128 stars per data point. The distribution by instrument is summarized in table~\ref{tab:res}. 
For all data we have a corresponding guide camera image FWHM value, as this is required input for the DIPSF code (see section~\ref{sec:anal} above).  In order to be useful for further analysis we also require a contemporaneous DIMM measurement. 

We query the DIMM database once for each night on which we have DIPSF data. Once we have downloaded the entire night's DIMM data it is straightforward to match them to our DIPSF exposures (and much faster than performing each individual match in SQL directly on the database). We average all DIMM readings taken during each DIPSF exposure to give the corresponding DIMM seeing for that exposure. An identical approach is taken with the guide camera readings. 
Unfortunately, the Cerro Manqui DIMM was not operated every night during the time period for which the DIPSF data was collected, as some of this period occurred after the GMT site testing program finished its main data collection. Therefore we do not have corresponding DIMM seeing values for the majority of DIPSF data points. After matching we have 876 concurrent data points from 62 nights ($\sim 10^5$stars). 
The breakdown by instrument is shown in table~\ref{tab:res}. 
Fig.~\ref{fig-time} shows examples of the variation of image quality throughout various nights. The DIPSF science image quality can be seen to track the DIMM and guide camera seeing values very well.

\begin{table}
\begin{center}
\caption{\label{tab:res} Summary of results: Number of fields, $n_F$, analyzed for each instrument, and average number of stars per field, $\bar{n_\star}$. The final column lists the number of fields with contemporaneous DIMM data, $n_{\mathrm cD}$.}
\scriptsize
\begin{tabular}{lrrrr}
\hline
Instrument & Field area (sq. arcmin) & $n_F$ & $\bar{n_{\star}}$ & $n_{\mathrm cD}$ \\
\hline
\hline
IMACS       & 635 & 1287& 459 & 199 \\
MagIC        & 6     & 2934 & 17 & 310 \\
LDSS3       & 54   & 1396 & 42 & 367 \\
\hline
\end{tabular}
\end{center}
\end{table}


\section{Results}
\label{sec-res}

\begin{figure*}[t]
\begin{center}
\plottwo{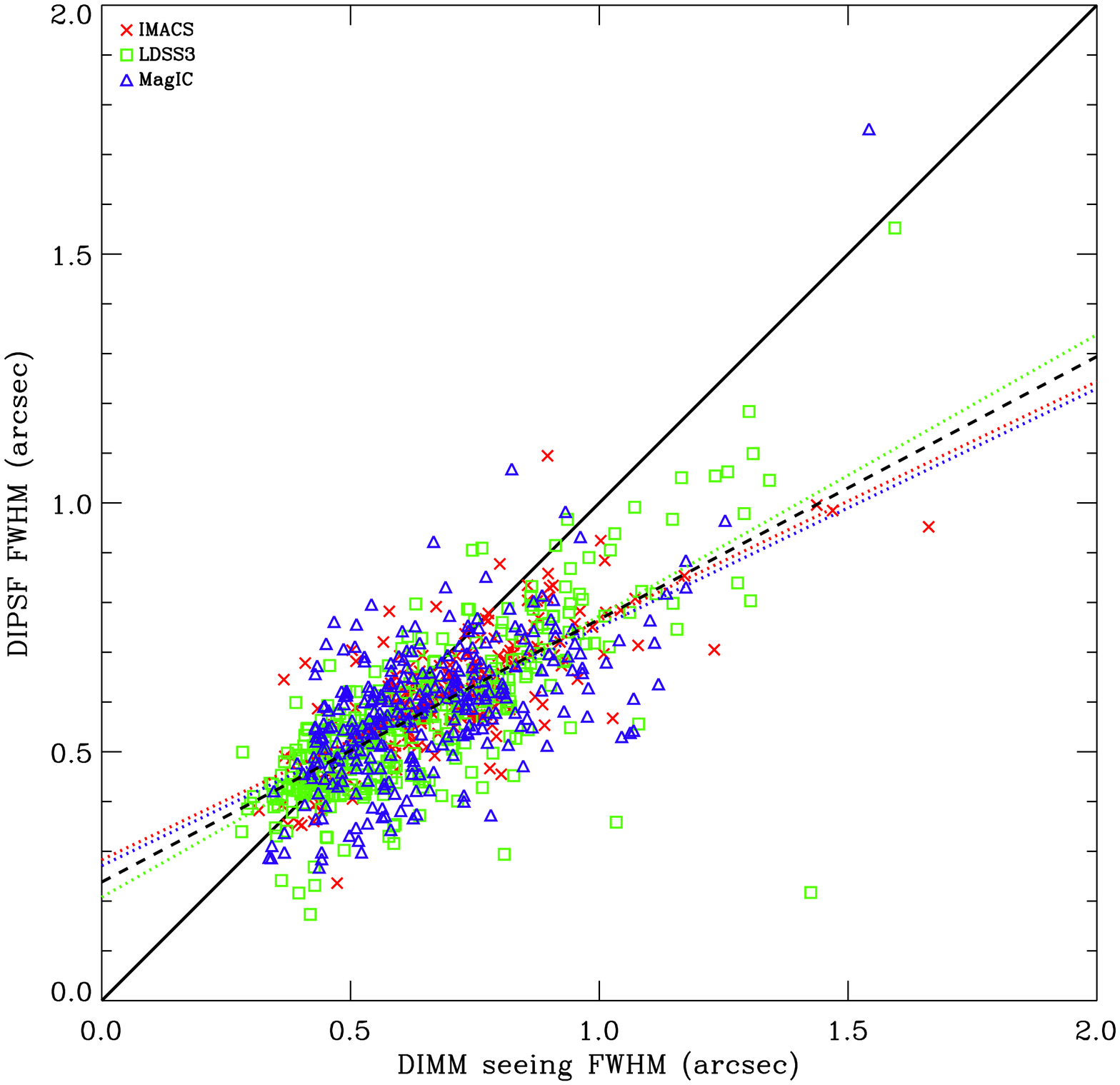}{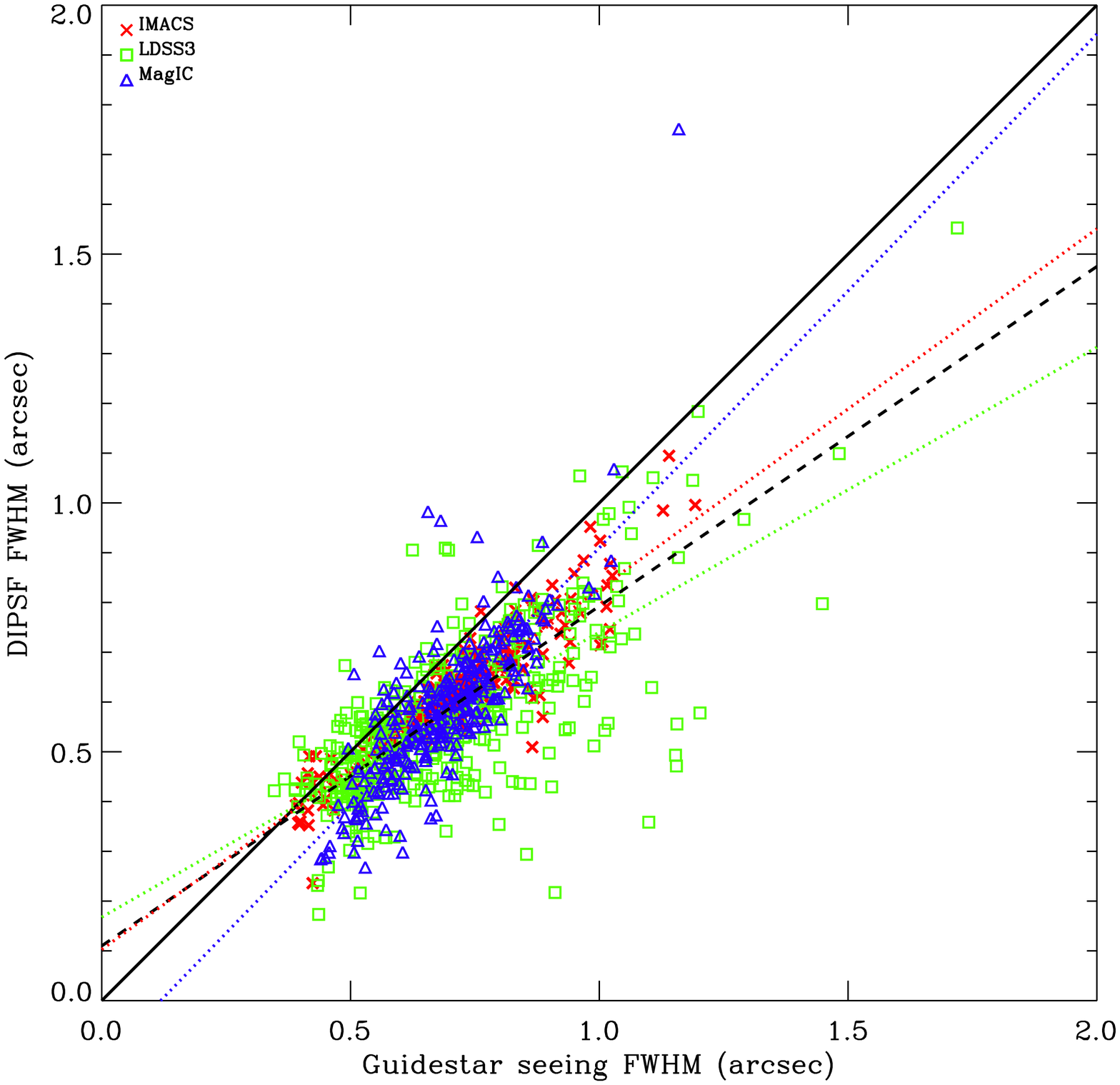}
\caption{\label{fig-ddg} FWHM as measured by DIPSF on science images against simultaneous DIMM seeing FWHM values (left) and guide camera seeing FWHM values (right). Red crosses denote IMACS data, green squares LDSS3 and blue triangles MagIC. 
The solid line on each panel shows 1:1 for reference. We also show best linear fits for the whole sample (black dashed line) 
and for each instrument individually (dotted lines in same colors as used for the symbols).}
\end{center}
\end{figure*}

In Fig.~\ref{fig-ddg} we plot the DIPSF seeing values against simultaneous readings for the DIMM seeing (left) and guide camera seeing (right). It is clear that the DIPSF and DIMM values are strongly correlated (Spearman's rank correlation coefficient $\rho = 0.71$; $p=0.00$). There is however a large dispersion, and a deviation from 1:1 (see left panel of the figure). 
The DIPSF and guide camera image qualities are also strongly correlated, and lie closer to 1:1, although with a larger dispersion (see right panel of the figure). The DIPSF and DIMM FWHM distributions are significantly different according to a two-sided Kolmogorov-Smirnov (K-S) test, which returns a cumulative difference $D=0.21$ between the two distributions, with a probability $p\sim10^{-17}$ of being drawn from the same underlying distribution. The difference is even more significant for DIPSF and guide camera image quality distributions, with a K-S test returning a cumulative difference difference $D=0.33$ with a probability $p\sim10^{-42}$.  

The median DIMM seeing for our matched dataset is 0\farcs625. 
The median DIPSF FWHM is 0\farcs575 and for the guide cameras it is 0\farcs688.  
Mean values and standard deviations are 
$\mathrm{FWHM_{DIMM}} = 0\farcs658\pm0\farcs205$ (skewness =1.04), 
$\mathrm{FWHM_{DIPSF}} = 0\farcs585\pm0\farcs150$ (skewness =1.24) and 
$\mathrm{FWHM_{guide}} = 0\farcs697\pm0\farcs165$ (skewness =0.923).

The Magellan image quality is better than the DIMM seeing 69\% of the time, and better than the guide camera image quality 90\% of the time. In poor seeing conditions (DIMM values $\ge$ 1\arcsec) Magellan does better than the DIMM 98\% of the time. The result is qualitatively as we would expect for a large telescope: In general the Magellan science cameras return an image quality that is better than the seeing as measured by the DIMMs. At good seeing we are limited by the telescope optics, and the two readings converge, while at poor seeing, the Magellan telescopes appear to do progressively better than would be expected from the DIMM seeing. 
The best linear fit to the entire dataset gives 
$\mathrm{FWHM_{DIPSF} = 0.53~FWHM_{DIMM} + 0.24}$. For the individual instruments this varies little (see Fig.~\ref{fig-ddg}, left panel), with LDSS3 following a slightly steeper relation than MagIC and IMACS. When comparing the DIPSF data to the guide camera FWHM there is a more marked difference between the instruments. 



\begin{figure*}[tb]
\vspace{5mm}
\begin{center}
\plottwo{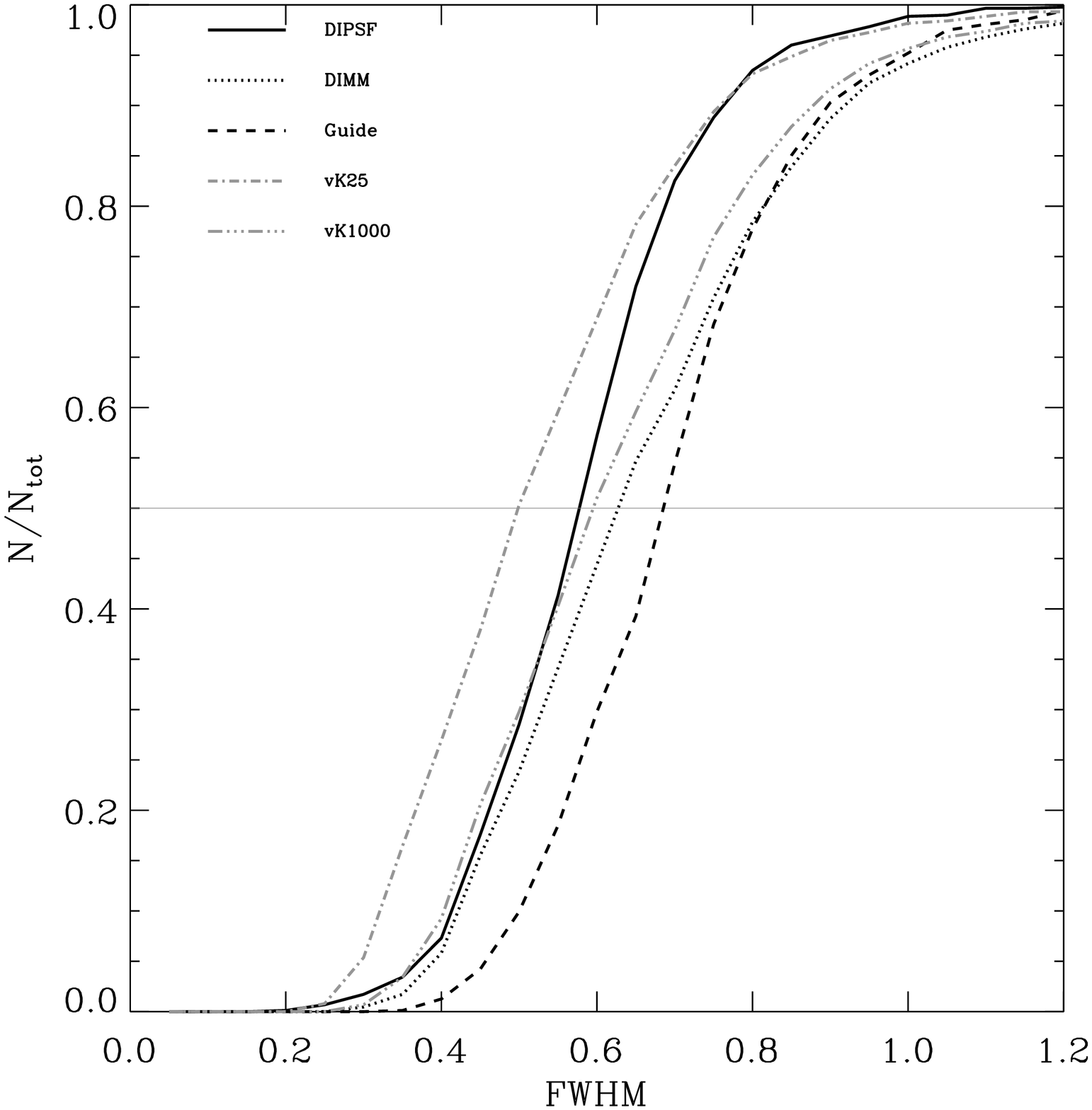}{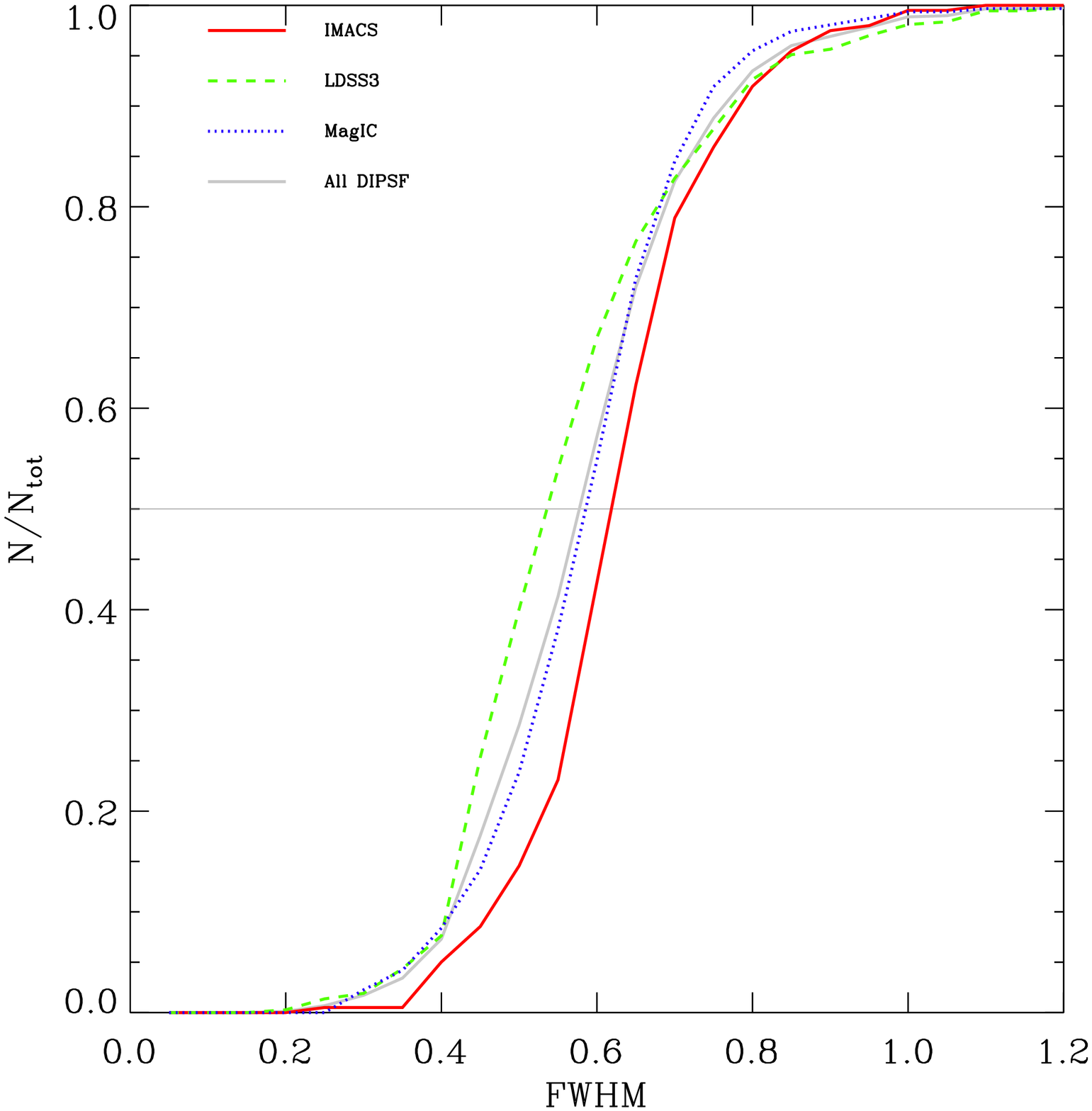}
\caption{\label{fig-hist} {\bf Left:} Cumulative distributions for the DIPSF, DIMM and guide camera FWHM measurements.  
\vk model predictions are shown for two cases:  $\mathcal{L}_0 = 25$~m (dot-dashed) and $\mathcal{L}_0 = 1000$~m (triple-dot-dashed). 
{\bf Right:} The instrumental breakdown of the DIPSF FWHM cumulative distribution (IMACS solid; LDSS3 dashed; MagIC dotted). 
To facilitate comparison with the left hand figure we show the DIPSF distribution for the complete dataset in grey.}
\end{center}
\end{figure*}

It is interesting to note that the IMACS DIPSF FWHM values do not get as low as those for the other instruments, in spite of having the largest number of measurements. The minimum IMACS DIPSF FWHM (excluding a single outlier at 0\farcs24) is at 0\farcs35. LDSS3 has a minimum FWHM at 0\farcs22 and 17 points that are lower than the IMACS minimum (excluding a single outlier at 0\farcs17). MagIC has a minimum FWHM of 0\farcs27 and 13 points that are below the IMACS minimum. This may be traceable to the averaging over a larger field of view, or inherent to the optics on the IMACS port of the telescope. 
The (S-H) wavefront sensors fitted to the instrument ports provide a measure of the optical aberations present during operation of the Magellan telescopes. These predict a modal Gaussian PSF FWHM of 0\farcs17.
This is sufficient to degrade a seeing of 0\farcs25 to 0\farcs 30. However, the S-H system on IMACS predict
a modal Gaussian PSF FWHM of 0\farcs30 FHWM. This is sufficient to degrade 0\farcs25 seeing to 0\farcs39. 
In Fig.~\ref{fig-hist} we plot the cumulative histograms the DIPSF, DIMM and guide camera FWHMs. On the right hand figure, we show the breakdown by instrument. We also present the cumulative histograms for various models with \Lo\ included. In good seeing the bulk of the DIPSF distribution tends toward that of the DIMM, consistent with the Kolmogorov model.  At the poor seeing end we see a migration of the DIPSF distribution toward lower FWHM values than the DIMM seeing. 
We find that a \vk model with an outer scale value of $\mathcal{L}_0=25$~m bounds the distributions.



\subsection{Outliers and observer bias}
Some nights result in DIPSF FWHM values that are significantly higher than the corresponding DIMM and guide camera FWHM values. In the vast majority of cases these are MagIC data points, and all of these are traceable to nights on which the observer is known to use out of focus images in order to increase the total detected source flux and thus signal-to-noise ratio. The remainder of the offending points arise from engineering runs. In all of these cases we have removed the entire night's data from the analysis. The numbers presented above and in table~\ref{tab:res} have these omissions taken into account. 

It is worth considering whether an additional bias is introduced by the observers themselves. 
Typically astronomers may drop an imaging program in favor of a backup spectroscopic program in the event of poor seeing, and this may artificially bias our results in favor of nights with good seeing. 
To test the robustness of our results we compare the average DIMM seeing for our subset of nights with that for all DIMM data taken over the same period. 
The median DIMM seeing for our data set is 0\farcs625 and that for the whole of the same time period (July 2007 -- December 2008) is 0\farcs675. 
T`hus we conclude that there is indeed an observer bias, and the median seeing reported above is slightly optimistic for the year in question. However, when we compare to other periods, we find that the DIMM seeing across our subsample is quite typical of long periods at Las Campanas. 
For example, from September 2005 -- September 2008 the median DIMM seeing was 0\farcs62~\citep{jet+10}; 
For the entire year 2007 the DIMM seeing was 0\farcs65~\citep{jet+08}; From April 2005 -- December 2006 the median DIMM seeing was 0\farcs65~\citep{jet+07}.
Thus we conclude that our estimate of Magellan image quality is robust for average conditions average, and that 2008 was an unusually poor year. 
It is worth noting that 2008 was a La Ni\~{n}a year, and one of the wettest in the southern Atacama for a decade. 

\subsection{Factors affecting the deviation of Magellan image quality from DIMM seeing}
It is clear from Fig.~\ref{fig-ddg} that the Magellan imaging FWHM is consistently better than the DIMM seeing and the guide camera FWHM. In particular we find that the Magellan telescopes do much better than the DIMM seeing when the seeing is poor.  In order to provide a more powerful analysis, let us define a ``seeing residual'' that is the difference between the DIMM seeing and the science image quality, \rse, such that a positive $r_s$ implies a better science image quality than the DIMM seeing. This gives us a direct measure of the improvement Magellan affords over the seeing for any given observations.  
We see no large-scale trend of seeing residual with telescope pointing and conclude that there are no major directional impediments to the image quality obtained at Magellan.

\begin{figure}[htbp]
\begin{center}
\plotone{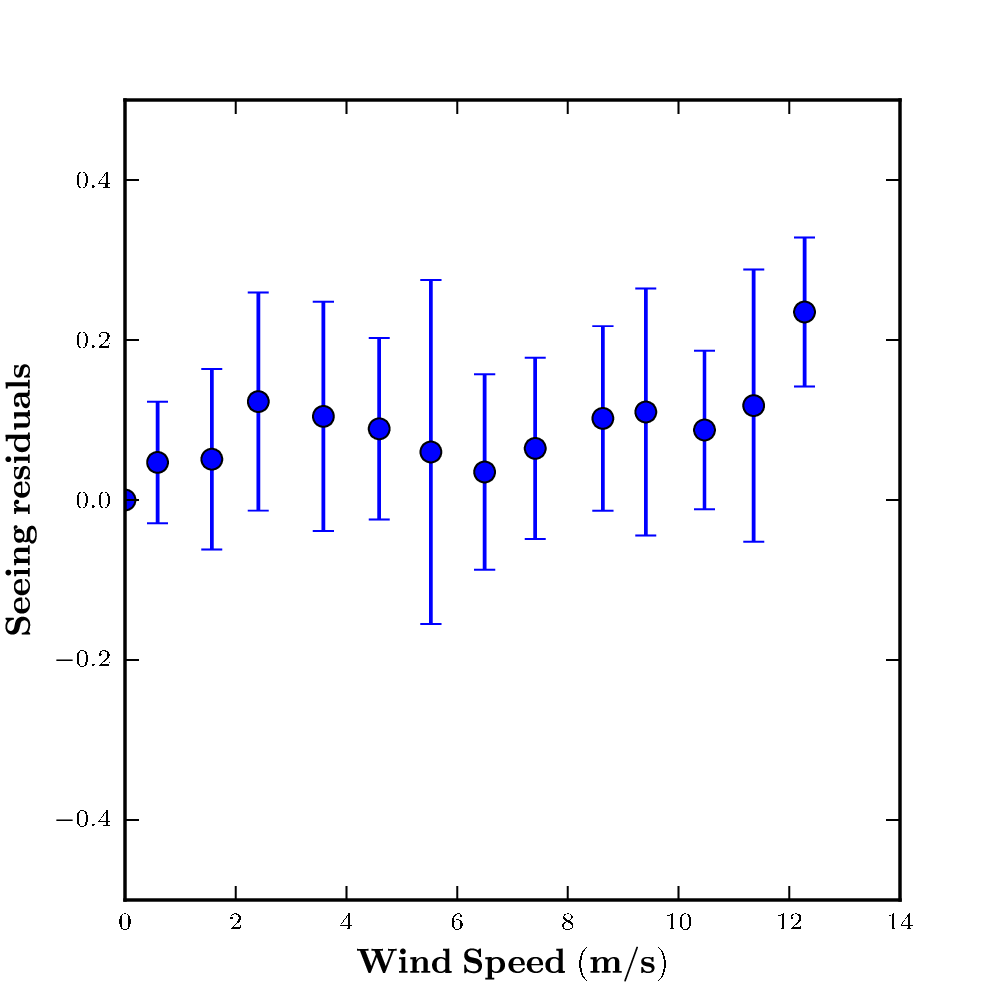}
\caption{\label{fig-windspd} Seeing residual, \rse\ with wind speed.}
\end{center}
\end{figure}

\begin{figure}[htbp]
\begin{center}
\plotone{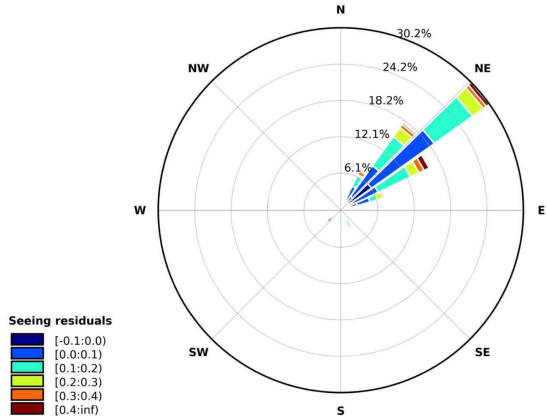}
\caption{\label{fig-windsee} Variation of seeing residual, \rse\ with wind direction in histogram format. The poorest DIMM seeing (red, orange) is dominated by east-northeasterly wind conditions and not by the prevailing northeasterly.}
\end{center}
\end{figure}

\begin{figure}[htbp]
\begin{center}
\plotone{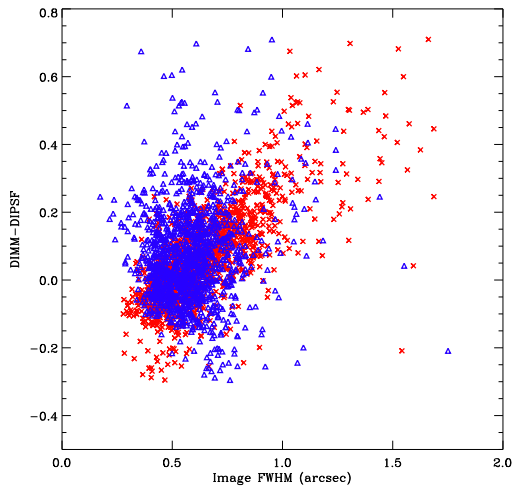}
\caption{\label{fig-resid} Seeing residual, \rse\ against DIPSF FWHM (black/blue triangles) and DIMM seeing FWHM (grey/red crosses). The residuals show a strong correlation with the DIMM seeing (Spearman's rank $\rho=0.69$ with a probability $p<0.005$ of occurring at random) and a much weaker correlation with DIPSF FWHM ($\rho=0.05$; $p=0.18$).}
\end{center}
\end{figure}

With the image quality database in place we are in a powerful position to begin exploring the effects of weather conditions on both the DIMM seeing and science image quality. It is known that the seeing correlates weakly with strong wind speeds~\citep{CoDR:GMT}. 
Figure~\ref{fig-windspd} illustrates the change of seeing residuals with wind speed. While there is always a weak tendency for $r_s>0$, this becomes strongest at the highest wind speeds (and worst seeing conditions). 
In Figure~\ref{fig-windsee} we illustrate the seeing residuals with wind vector in histogram form. No trend is seen with wind direction, beyond the fact that the prevailing wind direction at Las Campanas is North Easterly. 
Thus Magellan telescopes do relatively better than the DIMMs when the DIMM seeing is poor (as noted previously), and these situations often correspond to high wind speeds.  But is this due to a genuine improvement in the Magellan seeing, or just a worsening of the DIMM seeing? To help answer this we plot the seeing residual against both the DIPSF (blue triangles) and DIMM (red crosses) FWHM values in Fig.~\ref{fig-resid}. We clearly see that the residuals are more strongly correlated with the DIMM values (Spearman's rank $\rho=0.69$ with a probability $p<0.005$ of occurring at random) than they are with the DIPSF values ($\rho=0.05$; $p=0.18$). 
Thus the relative improvement in Magellan image quality in poor seeing conditions is found to be, to first order, a local effect at the DIMM towers. The DIMM method is insensitive to tracking errors and wind shake, and we suggest that the high winds may cause local turbulence through interaction with the clamshell dome. The dome opens almost due south (bearing 177), and so will interact with the prevailing wind upstream of the DIMM apertures.

\begin{figure*}[htbp]
\begin{center}
\plottwo{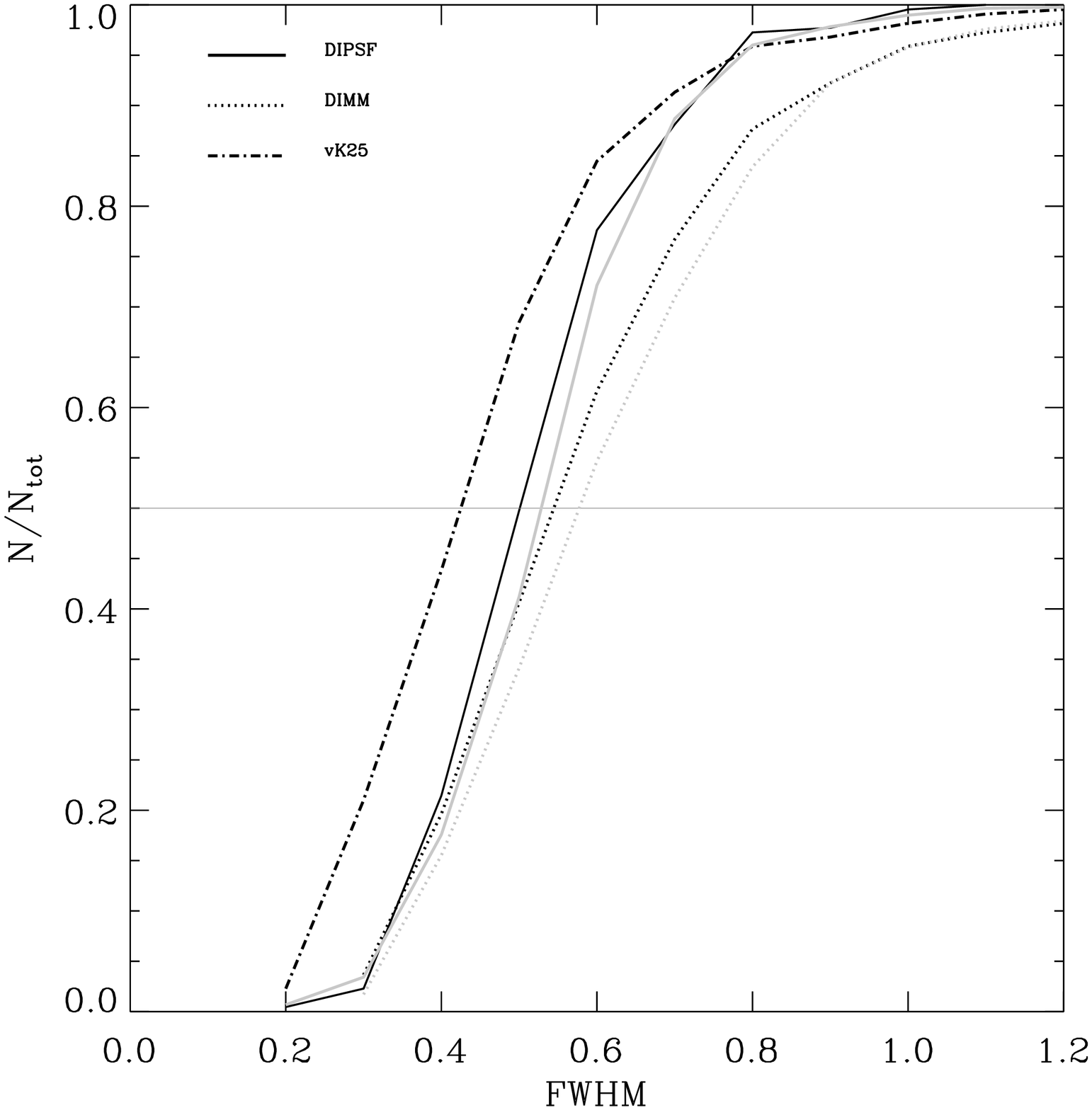}{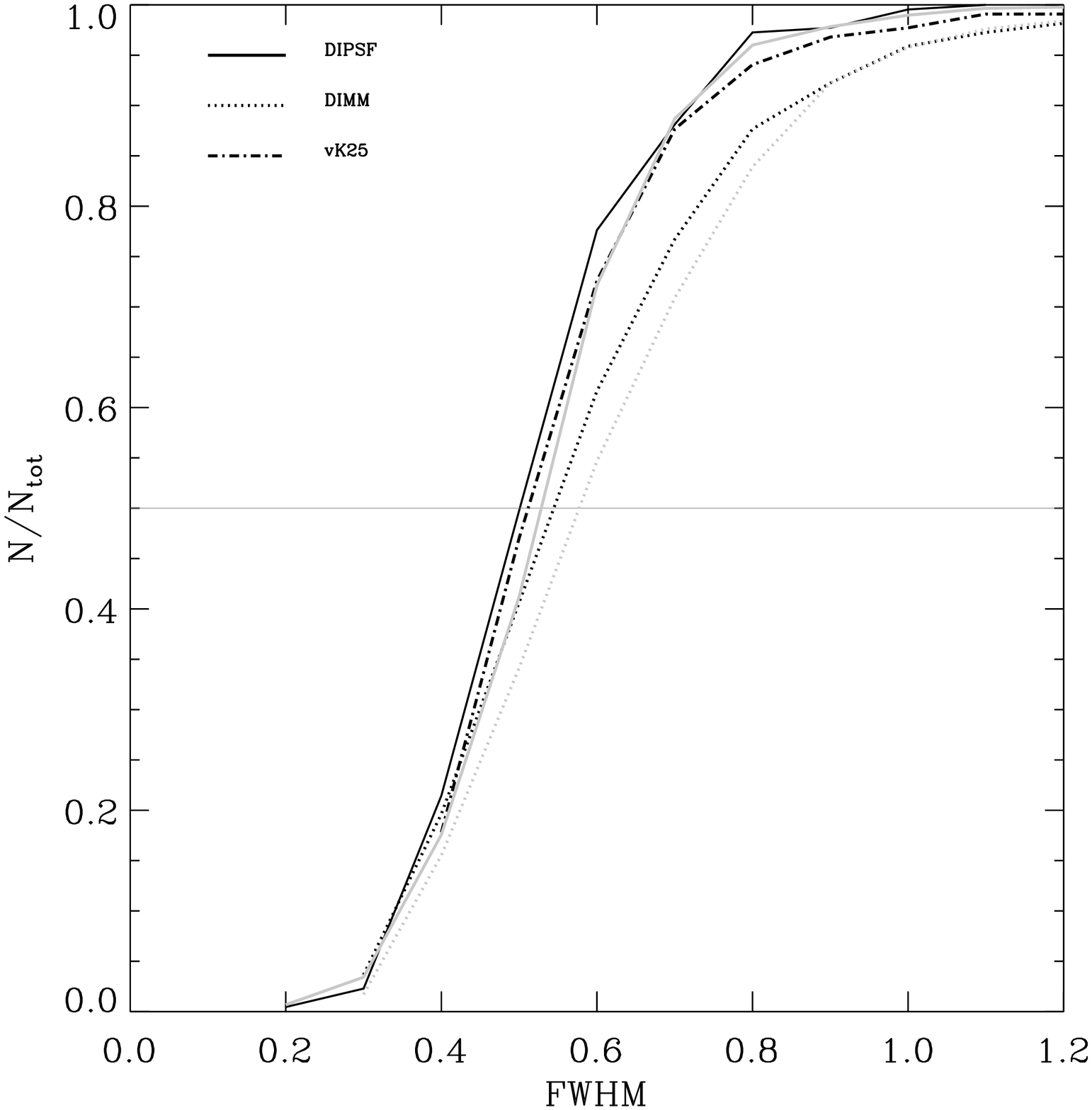}
\caption{\label{fig-CLoWind} The cumulative image quality distributions for the DIPSF (solid black) and DIMM (dotted black) data for those data points where wind speed $<10$~m~s$^{-1}$. The \vk model distribution with $\mathcal{L}_0=25$~m is shown by the dot-dashed black line. For reference, the original DIMM and DIPSF distributions for the full dataset (see Fig.~\ref{fig-hist}) are shown in grey. Removing the high wind speed data has the effect of improving the DIMM seeing (and to a lesser extent the DIPSF image quality), especially at poor seeing. Right: The same figure with 0\farcs3 added in quadrature.}
\end{center}
\end{figure*}

We now remove all datapoints for which the wind speed exceeds 10~m~s$^{-1}$.  This yields the cumulative image quality distributions shown in Fig.~\ref{fig-CLoWind}. 
Removing the high wind speed data primarily improves the DIMM distribution, with a smaller improvement seen in the DIPSF distribution. 
The median DIPSF image quality drops from 0\farcs575 to  0\farcs553, whereas the median DIMM image quality drops from 0\farcs625 to 0\farcs594.
The \vk model distributions are defined by the DIMM distribution and are therefore also affected.
We see from the figure that the DIPSF distribution still cuts across a range of \vk models. 
We note that an outer scale of $\mathcal{L}_0 \approx 25$~m (consistent with other sites) gives a conservative estimate of the image FWHM at seeing values above 0\farcs6. 
At low seeing values it seems likely that we are beginning to hit limitations imposed by the telescope optics. 
As discussed above (see foot of Section~\ref{sec-dataset}), the LDSS3 and MagIC ports are slightly better than the IMACS port, but in general we would expect even the best seeing conditions ($\sim0\farcs25$) to return image FWHM of $\sim 0\farcs3 - 0\farcs4$ on the instrument ports, and this is observed in Fig.~\ref{fig-CLoWind}. 



\subsection{Negative DIMM bias}
\label{sec-09}
As mentioned in section~\ref{sec-dimmbias}, we found our 2009 DIMM data to be affected by optical aberrations introduced by mis-collimation and the resulting loss of ability to achieve a proper focus.  Optical aberrations are expected to cause both negative and positive bias up to a ~20\%~\citep{tok+07}.  While positive biasing has been demonstrated experimentally before by the TMT site testing team~\citep{wang+07see}, no reports have been published of a negative bias.

\begin{figure}[htb]
\begin{center}
\plotone{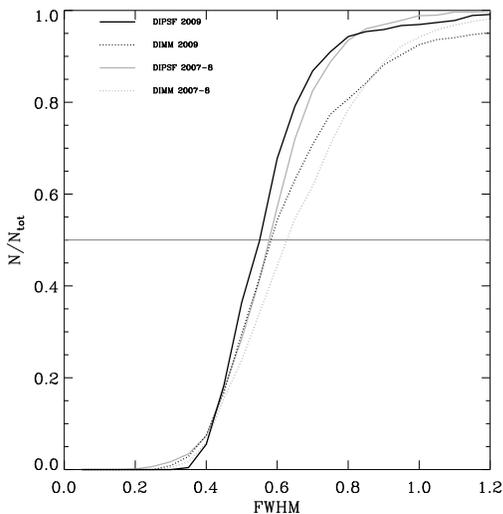}
\caption{\label{fig-hist09} Cumulative FWHM distribution for 2009 for DIPSF (solid black line) and DIMM (dotted black line).
The distribution for the data from July 2007 -- December 2008 (as used in the bulk of this paper and plotted in Fig.~\ref{fig-hist}) is shown in grey for reference.}
\end{center}
\end{figure}

The 2009 data set consists of 103 nights worth of data between January and October 2009. There are 456 DIMM-DIPSF matches over 28 nights. The 2009 dataset is therefore roughly half the size of the 2007--2008 dataset (876 points over 66 nights).
Figure~\ref{fig-hist09}  compares the 2009 DIPSF and DIMM data seeing distributions to those used in the above analysis.  We found that in 2009 the DIMM profile was biased towards better seeing in good conditions and worse seeing in bad conditions. As discussed in \cite{tok+07}, a negative bias can occur for high altitude turbulence under certain conditions of optical aberration.  In general at LCO, good seeing conditions occur when there is little ground layer seeing and the majority of the atmospheric turbulence occurs high up in the atmosphere.  In this case, it is possible for the DIMMs to show a negative bias, depending on the type of optical aberration present. Bad seeing conditions are more of a mix in terms of where the turbulence is located and thus may create more positive than negative biases.

\begin{figure*}[htb]
\begin{center}
\plottwo{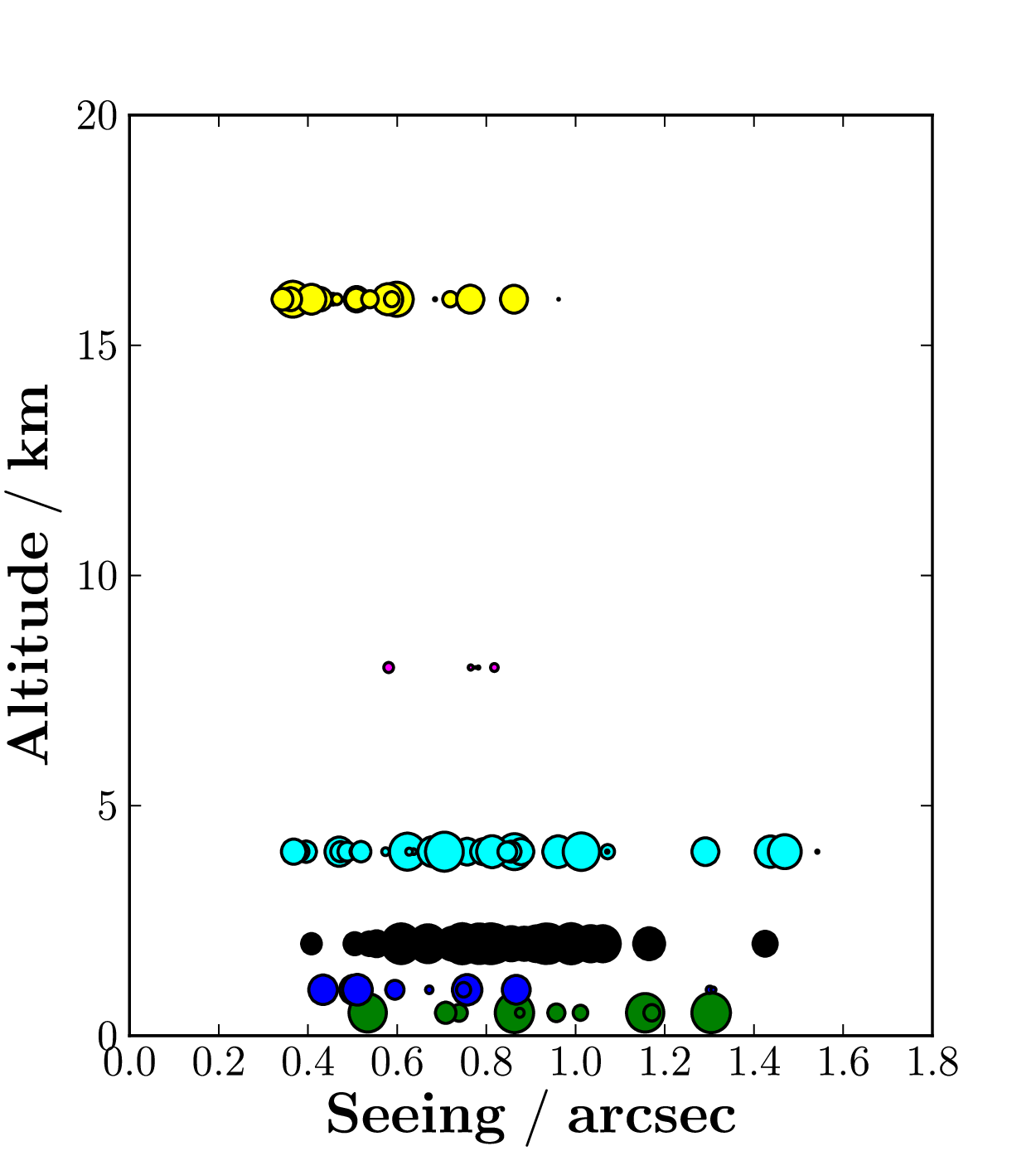}{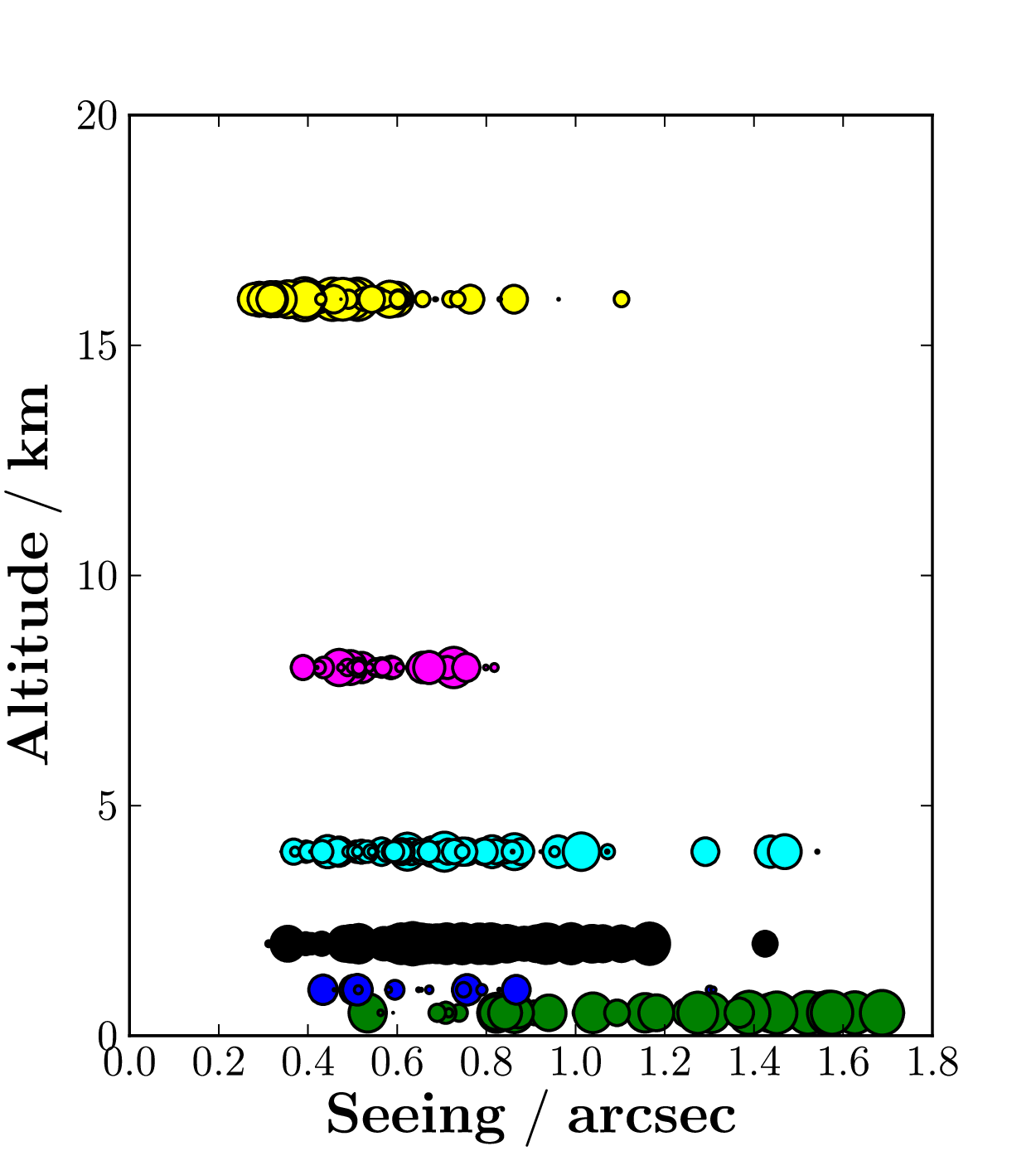}
\caption{\label{fig-lyrs} 
The altitude of maximum turbulence in the MASS free atmosphere $C_n^2(z)~dz$ profile as a function of seeing. 
The point size is scaled (by a function of the secondary maximum) to indicate how dominant the turbulence is in that layer. 
{\bf Left:} the 2007--2008 dataset used in the bulk of this paper. 
{\bf Right:} With the addition of the biased 2009 dataset (see section~\ref{sec-09}).}
\end{center}
\end{figure*}

We used the MASS $C_n^2(z)~dz$ profiles and DIMM seeing to explore the possibility of a correlation between the altitude of maximum turbulence and the seeing. Fig.~\ref{fig-lyrs} plots the altitude of maximum turbulence as a function of seeing: the point size is an indication of how dominant the turbulence is in that layer (i.e. how much larger the maximum is than the secondary maximum in the profile).  We can see that for the dataset including the biased 2009 data (right panel), the 16~km altitude bin clearly dominates in good seeing while this is less prominent in the unbiased dataset.  Furthermore, there is also more turbulence seen in the 8~km bin in the biased dataset. Thus the negative bias detected is  indeed associated with high altitude turbulence.

\subsection{Implications for observations in the NIR}
\begin{figure}[htbp]
\begin{center}
\plotone{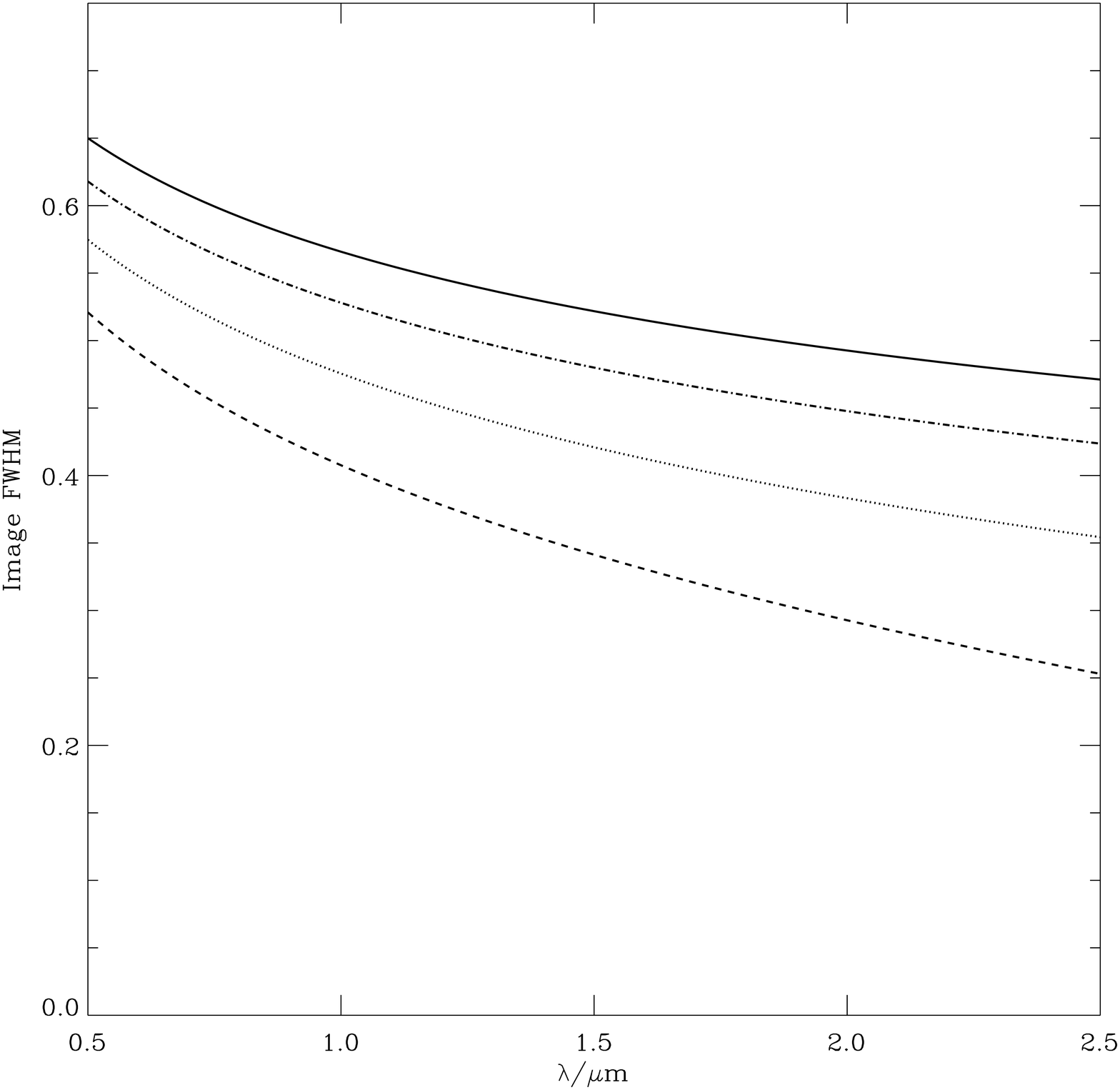}
\caption{\label{fig-see-lam} The expected mean average FWHM with wavelength $\lambda$ at Las Campanas Observatory.
The solid line indicates the Kolmogorov model, appropriate for small telescopes where the seeing is governed by $r_0$.
For telescopes that are an appreciable fraction of \Lo\ we expect an improvement over the Kolmogorov model due to the outer scale effect. 
We show the relationship for three $\mathcal{L}_0$ values: 25~m (dashed); 100~m (dotted) and 1000~m (dot-dashed).}
\end{center}
\end{figure}

\Lo\ is independent of wavelength and thus the gain in image quality for a large telescope as we move redward is somewhat greater than we would expect for the strict case of Kolmogorov turbulence -- see Fig.~\ref{fig-see-lam}. 
If we adopt an average seeing of 0\farcs65 and assuming $\mathcal{L}_0=25$~m we would estimate that for a large telescope the ideal image quality (assuming perfect optics) at 500~nm is 0\farcs52. 
In reality, we are degraded somewhat due to imperfect optics, as seen from the Shack-Hartmann system.
If we extrapolate further, we estimate that the ideal image quality for a large telescope should be 0\farcs41 at 1$\mu$m, 0\farcs29 at 2$\mu$m, and 0\farcs26 at 5$\mu$m.

\section{Conclusions}
We have presented a simple method for measuring the outer scale of turbulence that can be used at any observatory site that has a large aperture telescope used regularly for imaging, and a DIMM device for measuring the seeing. 
The Las Campanas Observatory site offers truly excellent seeing conditions. The effective ``seeing'' for the Magellan telescopes is considerably underestimated by the DIMMs. While the DIMMs give a median seeing of 0\farcs625, the median Magellan science image FWHM is 0\farcs575.
We find the dominant cause of this difference is local to the DIMM: high wind velocities introduce local turbulence, perhaps through interaction with the clamshell dome.
There is weak evidence for a directional influence on the DIMM image degradation at wind directions perpendicular to the orientation of the opening.
However, removing points with wind speeds $>10$~ms$^{-1}$ we find that the Magellan image quality is still significantly better than the DIMM seeing. From the cumulative distributions of the DIPSF and DIMM FWHMs we estimate that the site has an average turbulence outer scale of $\mathcal{L}_0\sim25$~m, similar to that measured at other sites.

We were unable to include our 2009 data due to a mis-collimation of the Cerro Manqui DIMM. 
This leads to both positive and negative biases in the DIMM data for the year, and is the first time that such negative biases have been recorded, although they are predicted. These negative biases occur when the upper layers of the atmosphere dominate the atmospheric turbulence profile.

\acknowledgments
This paper includes data gathered with the 6.5~meter Magellan Telescopes located at Las Campanas Observatory, Chile.
DF acknowledges the support of a Magellan Fellowship from  Astronomy Australia Limited, and administered by the Anglo-Australian Observatory.
We thank the anonymous referee for a constructive report that helped us clarify and improve aspects of the paper.
We gratefully acknowledge valuable discussions with Andrei Tokovinin, Paul Schechter, Mark Phillips, Povilas Palunas and Rebecca Sobel at various stages in the development of this work. In particular we thank Andrei Tokovinin for reviewing an early draft of this paper, and Paul Schechter for great assistance with DIPSF.
We are indebted to the Magellan Instrument Specialists, Jorge Bravo, Gabriel Martin, Victor Merino and Mauricio Navarrete for daily running of the code, and to Cesar Muena of the GMT site testing team for maintaining MASS-DIMM site monitors in a working and well-calibrated condition. We thank all the staff at Las Campanas Observatory for their enormous hospitality and camaraderie. 

\bibliographystyle{astron}
\bibliography{MagSee,tools,magellan,Seeing}

\end{document}